\documentstyle[aps,prl,twocolumn,psfig,epsf]{revtex}
\begin{document}
\draft
 
\twocolumn[
\title{Evidence of Microscopic-Scale Modifications in Optical Glasses 
Supporting Second Harmonic Generation}
 
\author{C. Cabrillo, G.J. Cuello, P. Garc{\'\i}a-Fern{\'a}ndez and
F.J. Bermejo}
\address{Instituto de Estructura de la Materia,
Serrano 123, Madrid E-28006, Spain}
 
\author{V. Pruneri, F. Samoggia and P. G. Kazansky}
\address{Optoelectronics Research Centre, University of
Southampton, \\ Southampton S017 IBJ, United Kingdom}
 
\author{S.M. Bennington}
\address{Rutherford Appleton Laboratory, Chilton, Didcot, Oxon,
OX11 0QX, United Kingdom}

\date{\today}
\maketitle
\mediumtext
\begin{abstract}
\leftskip 2.0truecm      
\rightskip -2.0truecm    
\indent
We explore, by means of inelastic neutron scattering, the extent of 
changes in dynamic correlations induced by thermal poling of 
vitreous $\rm SiO_2$. 
The measured vibrational density of states shows an excess of modes in 
certain frequency regions as well as a narrowing of the 100 meV peak. 
These findings indicate that such alterations cannot be ascribed to the 
appearance of new well defined vibrational modes, such as those coming 
from localized topological defects, but rather arises from an increase 
in ordering in the material as attested in a reduced spread of the 
inter-tetrahedral angles.
\end{abstract}
]
\narrowtext
The ability of optical glass fibers to generate visible (green) light 
via Second Harmonic Generation (SHG) of intense infrared pump light 
remained unexplained for a long time since the first experimental 
detection of the phenomenon in 1985 \cite{OSTE86}. 
Two main reasons made such phenomenon unexpected. 
Firstly, a fundamental principle derived on symmetry grounds for the 
second-order susceptibility tensor \cite{YARI84}, forbids SHG in an 
isotropic material, such as amorphous silica. 
Secondly, a well defined phase matching between the interacting waves is 
required to allow constructive interference leading to efficient SHG. 
Even if some symmetry breaking mechanism is assumed, the way in which 
phase matching waves was automatically forced appeared as almost a 
``miracle''. The puzzle was essentially solved five years later by Dianov 
{\it et al.} \cite{DIAN91} by invoking the emergence of a spatially 
modulated local dc field, $E_{0}$, which, via third-order 
nonlinearity ($\chi^{(3)}$), induces a spatially modulated second order 
nonlinearity ($\chi^{(2)} \propto \chi^{(3)} E_{0}$).  
This so called Electric Field Induced Second Harmonic Generation 
(EFISHG), which is responsible for the frequency doubling, relies on a 
third order nonlinearity which is finite in structurally isotropic 
materials.
The dc field is induced by photocurrents originated from ionization 
of color centers associated with impurities in the glass structure.
In the simplest model \cite{DIAN91},
quantum interference between two and one photon ionization leads to a 
modulation of the current, resulting in a phase matching 
(or more properly, quasi-phase matching).
Under such conditions the symmetry is broken by the local dc field
and no significant changes in the structure are needed.
 
Given the obvious technological interest of the phenomenon, a 
substantial effort has been focused towards the achievement of a 
permanent $\chi^{(2)}$ and to improve the SHG efficiency in glass. 
Nowadays, the so called thermal poling technique, which consists in 
applying high voltage ($\sim$ 5 kV) at elevated temperatures (${\rm 
\sim 270^{o}C-300^{o}C}$) is able to provide figures for SHG conversion 
comparable to those shown by some inorganic crystals \cite{Mye91}. 
It is also possible to implement quasi-phase matching efficiently in 
periodically thermally poled fibers \cite{Pru97}.
 
Contrary to what could be expected, there is a lack of studies carried 
out by direct means on structural and/or dynamical alterations 
associated with poling. 
Most are either macroscopic \cite{Miz88} or mesoscopic 
\cite{Mar95_Kaz96} in nature, rendering scarce information of the 
underlying microscopic physics. 
Up to now, probably the most direct investigation of the phenomenon 
comes from Raman spectroscopy studies \cite{Gab87,Kam90,Ale94}. 
In fact, significant changes between treated
(photoinduced) and untreated fibers were observed at rather well
defined frequencies and associated with changes in the density of
topological ``defects'' in the ${\rm SiO_2}$ glass structure. 
Such structural modifications were assigned as the main microscopic 
symmetry-breaking mechanism leading to SHG \cite{Gab87,Sto87}. 
A direct interpretation of the Raman data is, however, hampered by the 
need of previous knowledge of the possible structural alterations in 
order to properly assess the concomitant variations in the intricate 
couplings between the optical waves and electronic clouds governing the 
signal intensities. 
As a matter of fact, the aforementioned defects model cannot completely 
explain all the experimental results \cite{Miz88}. 
In spite of this, however, it is sometimes presented in the literature 
as the way of explaining the phenomenon.
 
Our aim here is to investigate the microscopic mechanisms leading to the 
appearance of the second-order non-linear optical response in poled 
glass, using for the purpose a beam of thermal neutrons as a probe which 
directly couples to the nuclei.
We thus carried out a set of inelastic neutron scattering (INS) 
measurements of the generalized frequency spectrum (also referred as 
vibrational density of states) covering the frequency range where 
previous reports from optical spectroscopies found measurable 
differences between untreated and poled samples. 
 
We have chosen to work with thermally poled samples (nowadays probably
the most promising poling technic in regarding possible technological
applications). The samples used in the experiments were made of 
electrically-fused quartz (Infrasil from Heraeus) with dimensions 
$40\times 40 \times 0.1$ mm. The thickness ($\sim 100 \mu$m) was 
chosen to be the smallest compatible with the polishing in order to 
increase the ratio poled volume {\it versus} unpoled volume. 
In fact, after thermal poling \cite{Mye91}, the nonlinearity is located 
in a region $\sim 5 \mu$m deep under the anodic surface independently of 
the sample thickness, so that it is evident the convenience of using 
thin samples. 
A ratio of poled/unpoled material of $\sim 5\%$ was estimated. 
The thermal poling was carried out at ${\rm 275^oC}$ with an applied dc 
voltage of $\sim 5$ kV for 40 minutes. 
The poled sample consisted in fifteen plates stacked together. 
For comparison a similar stack of unpoled samples was used.
 
The INS experiments were carried out on the MARI chopper spectrometer at 
the ISIS pulsed neutron source at the Rutherford Appleton Laboratory, 
Oxfordshire (UK).
The spectrometer operates in direct geometry mode, that is the 
wavelength (energy) of the incident neutron pulse is monochromatized by 
a set of three rotating devices (choppers) which select neutrons having 
the required velocity. 
The energy and momentum analysis of the outgoing neutrons, after
scattering from the sample, is carried out by means of time-of-flight
measurements once neutrons are detected by a large array of $\rm ^3 He$
counters which span a large angular range. 
Two different incident energies were used (220 meV and 150 meV) which 
enabled the coverage of kinematic ranges with different resolution in 
energy-transfers (the achieved resolution varies as a smooth function of 
$\Delta E/E$, being typically of a few per cent).
In all the experiments the samples were kept at 20 K in order to reduce 
the multiple-phonon contribution to the spectrum, a quantity which 
becomes substantial in measurements of this kind performed at room 
temperature. 
The quantity which is directly accessible to a neutron scattering 
experiment is the double-differential (${\rm d}^2\sigma/{\rm d}\Omega 
{\rm d}E$) scattering cross-section given in units of barns meV$^{-1}$ 
Sterad$^{-1}$.
For a macroscopically isotropic material such as a glass, the quantity 
of utmost importance to specify the dynamics is the $Z(\omega)$ spectral 
frequency distribution (or vibrational density of states). 
For samples where the dominant neutron-nucleus scattering is of 
incoherent nature, derivation of $Z(\omega)$ from the measured 
cross-section only involves a few corrections. 
In the case of silica, where scattering from Si and O nuclei gives rise 
to coherent (interference) effects, the derivation of $Z(\omega)$ from 
the measured data becomes more elaborated. 
A number of schemes for averaging-out the coherent
effects have been developed and are known to lead to reliable
estimates of the spectrum. In our particular case we have adopted
a procedure described in detail in Ref. \cite{Daw96} which is based
on a phonon expansion of the dynamic structure factor $S(Q,E) \propto 
{\rm d}^2\sigma/{\rm d}\Omega {\rm d}E$ and consists in an iterative 
scheme designed to correct a number of instrumental and 
sample-dependent (multiphonon) effects.
On the other hand, having a coherently-scattering sample is here an 
additional bonus rather than a nuisance. 
In fact, the single-differential scattering cross-section ${\rm d}\sigma
/{\rm d}\Omega$ derived by integration over energy-transfers of the 
measured response is directly proportional to the static structure 
factor $S(Q)$ and therefore some hints of structural modifications could 
possibly be derived from the same set of experimental data. 
Besides, the wavevector dependence of $S(Q,E)$ at a given energy 
transfer provides direct information about the phase relationships of 
the motion of specific atomic pairs \cite{Car85_Ara91}, since the 
Fourier transform of $S(Q, E =$ ctant.) is a dynamical pair correlation 
function $D_{\rm d}(r,E =$ ctant.) which serves to identify the atoms 
taking part in motions at such frequency as well as their relative 
phases.
 
\begin{figure}
\centerline{\psfig{figure=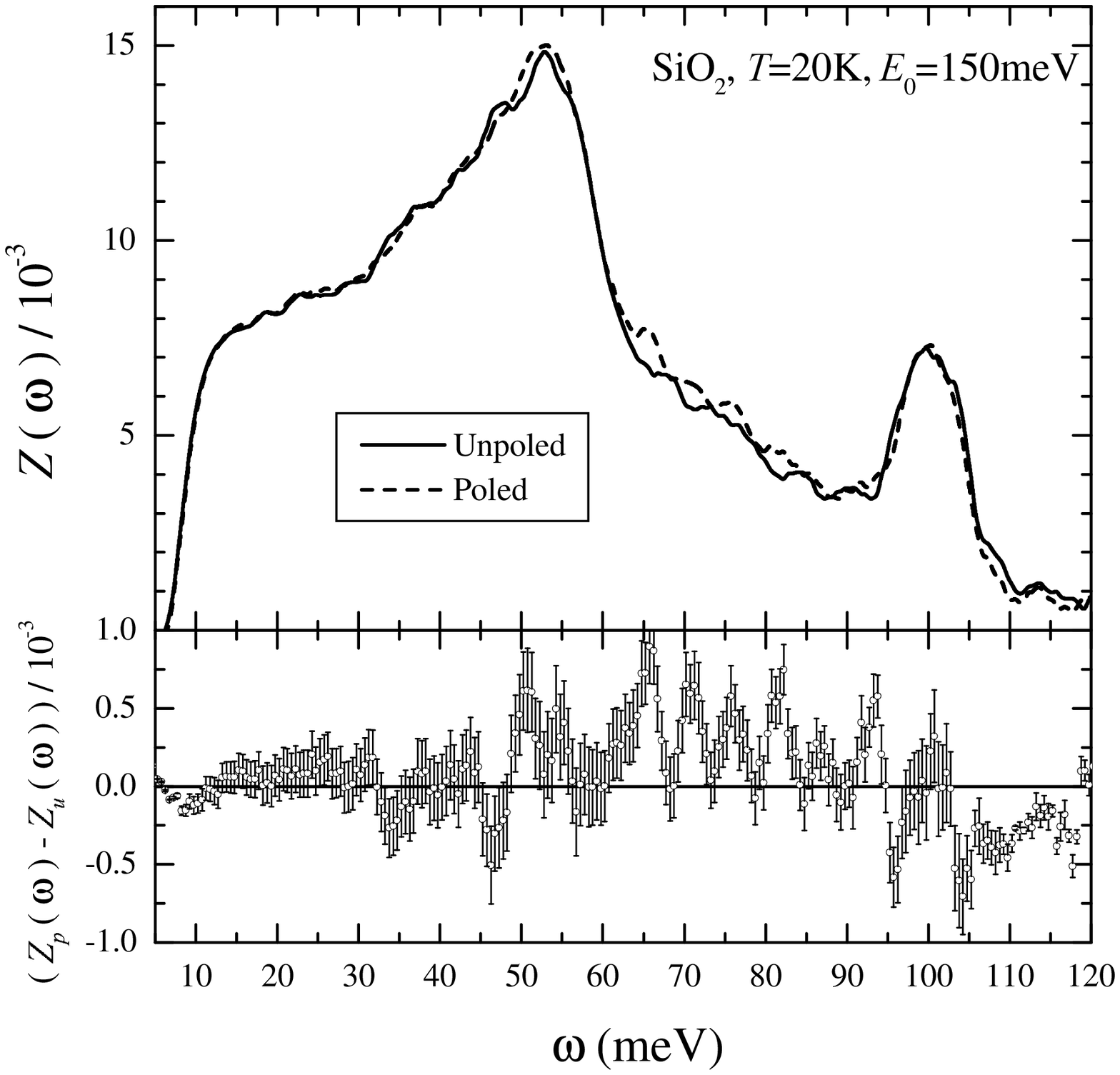,width=8.5cm,clip=}}
\caption{The $Z(\omega)$ frequency distribution of both, the virgin
sample (solid line) and the treated one (dashed line) as well as the
difference between them.}
\label{zeta}
\end{figure}

Figure \ref{zeta} displays a comparison between spectral distributions 
for the virgin (untreated) and poled samples up to a frequency of 120 
meV ($\rm \approx 960 \; cm^{-1}$). 
As stated above, such a range comprises the one where the Raman studies 
reported changes induced by poling. 
A measurable difference in $Z(\omega)$ between the poled and unpoled 
glasses is readily seen for frequencies about 53 meV ($\rm \approx 424 
\; cm^{-1}$), a band comprising 60 meV ($\rm \approx 480 \; cm^{-1}$) 
$\leq E \leq$ 95 meV ($\rm \approx 760 \; cm^{-1}$) and the peak at 100 
meV ($\rm \approx 800 \; cm^{-1}$). 
Notice that only 5\% of the sample was really poled and therefore the 
changes are correspondingly scaled down. 
As can be seen from the lower frame of the graph, the differences are 
outside error bars and this supports the statistical significance of 
such findings.
The measurements carried out using a larger incident energy (220 meV) 
confirmed the above mentioned changes and enabled the exploration of the 
spectral region extending up to 180 meV ($\rm \approx 1440 \; cm^{-1}$), 
which comprises the strong double-peak structure \cite{Car85_Ara91} with 
maxima at about 135 meV ($\rm \approx 1080 \; cm^{-1}$) and 150 meV
($\rm \approx 1200 \; cm^{-1}$). Feeble changes were found within that 
range of frequencies. 
However, the lower resolution and statistics prevents to discriminate a 
reasonably clear difference and therefore, our ensuing discussion will 
only concern alterations below 110 meV.

To compare the present results with the previous Raman studies, an 
analysis of the $Z(\omega)$ as customary in the Raman literature 
\cite{Gal91} seems in order. 
It is based on the band-edge normal modes of a simple analytical model, 
{\it i.e.}, a continuous random network of $\rm Si_2(O_{1/2})_4$
tetrahedra with only central forces \cite{Sen77}.
The stronger peak at $\approx$ 53 meV corresponds to the so called 
$\omega_1$ mode and is portrayed as a vibration where the oxygens 
undergo a symmetric stretch whereas the Si atoms are still. 
The well defined feature at $\approx$ 100 meV would then correspond to 
the $\omega_3$ mode where all atoms move and the pair of higher 
frequency peaks are assigned as arising from LO$-$TO splitting, the 
higher frequency (150 meV) being identified with the $\omega_4$ 
vibration.
The model is supplemented by two ``defect'' modes associated with 
threefold and fourfold planar rings of Si$-$O bonds in order to 
explain the so called D$_1$ and D$_2$ narrow Raman lines.
 
In spite of giving a neat interpretation of many of the Raman spectrum 
features, the above picture seems fairly oversimplified for the far 
broader shapes of the INS spectra (which contrary to the Raman do not 
depend upon intricate coupling-coefficients obeying selection rules). 
In particular, and central to our case, the D$_1$ and D$_2$ lines cannot 
be easily identified with clear features in the INS spectra. 
Instead, a broad irregular structure appears between 60 meV and 85 meV 
as shown in Figure~\ref{zeta} (also present in previous experiments 
\cite{Ara92}). From computer simulations using realistic potentials 
for the interparticle interactions \cite{Tar97}, we know that in this 
frequency range, where the material response is modified by poling, the 
vibrational dynamics still shows substantial ``collective" character, 
even if what is left at such high frequencies can only be considered as 
remnants of the long-wavelength phonons which become at these scales 
heavily damped and hybridized with vibrations of ``optical" character. 
These observations are in contrast with the highly localized modes of 
the ring-like topological ``defects'' model. From this new perspective, 
no structural changes are necessarily implied as the local dc fields 
are expected to be very high (near the dielectric breakdown value) 
and the driving of the electronic clouds on the nuclei could well 
affect their mesh of vibrational dynamics with negligible 
perturbations on their equilibrium positions.
 
In addition, two more spectral features are clearly affected by the 
poling, {\it i.e.}, the increase in the intensity of the broad peak 
centered at 53 meV and the narrowing of the 100 meV peak. 
It is not clear whether the former has been detected in the Raman 
experiments on photoinduced poling. 
In Ref. \cite{Kam90} very clear periodic changes in the intensity at 53 
meV were observed when a micro Raman was scanned along the fiber but 
they completely disappeared after averaging. 
This suggests a modulation of the Raman signal due to changes in the 
optical wave-electron coupling induced by the dc grating rather than 
changes in the vibrational density of states itself. 
For instance, periodic birefringence caused by the grating would affect 
the Raman intensity at 53 meV as it is strongly polarization dependent. 
Anyhow, our INS spectra are free from this uncertainty and certainly an 
increase in the 53 meV intensity is apparent. 
A recent analysis \cite{Mar96} shows how disorder alters the peak 
heights of the density of modes owing to coupling between close enough 
modes. Whether this gives rise to an increase or a decrease 
depends on the specific case. The observed enhancement 
of the 50 meV peak could then signal a change in order.
Given the high anisotropic nature of the poling process, 
one would expect an increase in ordering. Such a picture is 
strongly supported by the narrowing of the 100 meV peak 
(not reported in the Raman experiments). The fact that
ordering should produce narrowing is not only physically 
very plausible, but is also supported by calculations. Indeed, 
the results of Ref. \cite{Mar96} predict such an effect too. 
More explicitly, Bethe lattice calculations demonstrate that the main 
effect of a reduction on the spread of the inter-tetrahedral angle 
distribution is a narrowing of the 100 meV peak \cite{Gal91}, precisely 
the most intuitive kind of ordering that would be expected from the 
thermal poling process. It remains an open question whether such 
ordering is a significant contribution to the measured optical 
second order nonlinearity.
 
In summary, inelastic neutron scattering is capable of monitoring the 
alterations occurring in fused silica at a microscopic level after the 
material is submitted to thermal poling. From the present data 
(the vibrational density of states) two main conclusions may be inferred. 
Firstly, the modifications in a frequency range
encompassing the so called D$_1$ and D$_2$ Raman lines are consistent 
with dynamical alterations involving rather large groups of atoms. 
This finding is in contrast with the highly localized features expected 
from a creation of three-fold and four-fold ring-like ``defects'' 
sometimes assumed as possible cause of the second order nonlinear 
optical response. Secondly, alterations around 53 meV and particularly 
at 100 meV strongly suggest a higher order in the treated material 
implying a reduced spread in the inter-tetrahedral angle distribution. 
Further studies are needed to clarify whether these changes can justify 
the value of second-order nonlinearities measured after poling.
 
\section{Acknowledgments}
Work supported in part by grants No. TIC95-0563-C05-03, No. PB96-00819, 
CICYT, Spain, and Comunidad de Madrid 06T/039/96. 
V. Pruneri and F. Samoggia acknowledge Pirelli Cavi (Italy) for their 
fellowship and studentship respectively.

\end{document}